\documentclass[11pt]{article}
\usepackage{graphicx}

\newsavebox{\hflrar}
\sbox{\hflrar}{\makebox[0pt][l]
{${\scriptstyle \leftharpoonup}$}{${\scriptstyle \rightharpoonup}$}}

\def \to {\rightarrow}

\begin{document}
\begin{center}
{\Large\bf Transverse Momentum Dependent Light-Cone Wave Function of $B$-Meson
and Relation to the Momentum Integrated One} \vskip 10mm
J.P. Ma and Q. Wang    \\
{\small {\it Institute of Theoretical Physics, Academia Sinica,
Beijing 100080, China }} \\
\end{center}

\vskip 1cm
\begin{abstract}
A direct generalization of
the transverse momentum integrated(TMI) light-cone wave function to define
a transverse momentum dependent(TMD) light-cone wave function
will cause light-cone singularities and they
spoil TMD factorization. We motivate a definition in which the light-cone
singularities are regularized with non-light like Wilson lines. The defined
TMD light-cone wave function has some interesting relations to the corresponding
TMI one.
When the transverse momentum is very large, the TMD light-cone wave function
is determined perturbatively in term of the TMI one. In the impact $b$-space
with a small $b$,
the TMD light-cone wave function can be factorized in terms of
the TMI one. In this letter we study these
relations. By-products of our study are the renormalization evolution
of the TMI light-cone wave function and the Collins-Soper equation of
the TMD light-cone wave function, the later will be useful for resumming
Sudakov logarithms.
\vskip 5mm
\noindent
\end{abstract}
\vskip 1cm
\par
Exclusive B-decays play an important role for testing the standard model
and seeking for new physics. Experimentally they are studied intensively.
Theoretically, there are two approaches of QCD
factorization for studying
these decays. One is based on the collinear factorization\cite{BBNS}, in which
the transverse momenta of partons in a B-meson are integrated out and their
effect at leading twist is neglected. The collinear factorization
has been  proposed for other exclusive processes for long time\cite{BL}.
Another one is based on the transverse
momentum dependent(TMD)
factorization\cite{KT1},
where one takes the transverse momenta of partons into account at leading
twist by meaning of TMD light-cone wave function.
The advantage of the TMD factorization is that it may eliminate
end-point singularities in collinear factorization\cite{EPS}
and some higher-twist
effects are included. The knowledge of the TMD light-cone wave function
will provide a 3-dimensional picture of a $B$-meson bound state.
However, it is not clear how to define
the TMD light-cone wave function in a consistent way to perform a TMD factorization
because of light-cone singularities\cite{Col1}.
\par
In the collinear factorization the light-cone wave function for a $B$-meson
moving in the $z$-direction with the four velocity $v$
is defined as\cite{LCB}:
\begin{equation}
\Phi_+(k^+,\mu) =\int \frac{d z^- }{2\pi}  e^{ik^+z^-}
\langle 0 \vert \bar q(z^- n) L_n^\dagger (\infty, z^-n)
  \gamma^+ \gamma_5 L_n (\infty,0) h(0) \vert \bar B(v) \rangle,
\end{equation}
where we used the light-cone coordinate system, in which a vector
$a^\mu$
is expressed as $a^\mu = (a^+, a^-, \vec a_\perp) = ((a^0+a^3)/\sqrt{2},
(a^0-a^3)/\sqrt{2}, a^1, a^2)$ and $a_\perp^2 =(a^1)^2+(a^2)^2$.
The field $h(x)$ is the field for
$b$-quark in the heavy quark effective theory(HQET) and $q(x)$ is
the Dirac field for a light quark in QCD.
The gauge link is $L_n$ defined with the light-cone vector $n^\mu =(0,1,0,0)$
as:
\begin{equation}
L_n (\infty, z) = P \exp \left ( -i g_s \int_{0} ^{\infty} d\lambda
     n\cdot G (\lambda n + z ) \right ) .
\end{equation}
In the definition the transverse momentum of $\bar q$ is integrated, resulting
in that the field $\bar q$ and $h$ are only separated in the light-cone direction.
We will call $\Phi_+(k^+,\mu)$ as transverse momentum integrated(TMI)
light-cone wave function.
A direct generalization of Eq.(1) by undoing the integration to define
the TMD light-cone wave function causes serious problems because
it has
light-cone singularities, if quarks emit gluons carrying
momenta which are vanishing small in the $+$-direction but large
in other directions.
Similar problems also appear
in defining TMD parton distributions and fragmentation functions
for inclusive processes.
It has been shown that one can take in the definition
the gauge link in the direction off the
light-cone direction $n$ and the factorization of inclusive processes
can be done without light-cone singularities\cite{CS,CSS,JMY}.
In this letter
we propose to define  the TMD light-cone wave function
with gauge links slightly off the light-cone and study
its relation to the TMI light-cone wave function.
\par
We introduce a vector $u^\mu=(u^+,u^-,0,0)$ and define the TMD light-cone wave function
in the limit $u^+ << u^-$:
\begin{eqnarray}
 \phi_+(k^+, k_\perp,\zeta, \mu) = \int \frac{ d z^- }{2\pi}
  \frac {d^2 z_\perp}{(2\pi )^2}  e^{ik^+z^- - i \vec z_\perp\cdot \vec k_\perp}
 \langle 0 \vert \bar q(z) L_u^\dagger (\infty, z)
  \gamma^+ \gamma_5 L_u (\infty,0) h(0) \vert \bar B(v) \rangle\vert_{z^+=0},
\end{eqnarray}
the gauge link $L_u$ is defined by replacing the vector $n$ with $u$ in $L_n$.
This definition has not the mentioned light-cone singularity, but it has an extra
dependence
on the momentum $k^+$ through the variable
$\zeta^2 = 4(u\cdot k)^2/u^2$. This extra dependence is useful.  The evolution
in $\zeta$ is controlled by the Collins-Soper equation\cite{CS} which
leads to the so-called CSS resummation formalism\cite{CSS}, and it
will be derived here.
The evolution with the renormalization scale $\mu$ is simple:
\begin{equation}
\mu \frac{\partial \phi_+(k^+, k_\perp,\zeta, \mu) }{\partial \mu }
 = (\gamma_q +\gamma_Q) \phi_+(k^+, k_\perp,\zeta, \mu),
\end{equation}
where $\gamma_q$ and $\gamma_Q$ is the anomalous dimension of the light quark field
$q$ and the heavy quark field $h$ in the axial gauge $u\cdot G=0$, respectively.
\par
If one integrates $k_\perp$ in $\phi_+(k^+, k_\perp,\zeta, \mu)$,
the TMD light-cone wave function will in general not reduce to the TMI
one. The reason is that the integration over $k_\perp$
in Eq.(1) has
ultraviolet divergences, a renormalization is needed. Hence the integration
over the transverse momentum in Eq.(1) is done in $d-2$ dimension, if one uses
$d$-dimensional regularization, and then a UV subtraction is performed.
In contrast, the transverse momentum $k_\perp$
in $\phi_+(k^+, k_\perp,\zeta, \mu)$ is in the physical space with $d=4$,
ultraviolet divergences will be generated if one integrates over $k_\perp$
and they are not subtracted in Eq.(3) because it is a distribution of $k_\perp$.
Therefore, one can not simply relate $\phi_+(k^+, k_\perp,\zeta, \mu)$ by integrating
$k_\perp$ to the TMI light-cone wave function in Eq.(1).
However, the TMD light-cone wave function has some interesting relations
to the TMI one.
If the transverse momentum $k_\perp$ carried by the parton $\bar q$
is much larger than the soft scale $\Lambda_{QCD}$,
the $b$-quark as a parton will also carry large transverse momentum
because the momentum conservation. This can only happen if hard gluons
are exchanged between the two partons and the exchange can be studied
with perturbative QCD. Without the exchange of hard gluons, one expects
that the partons will carry $k_\perp$ with a typical value at order of
$\Lambda_{QCD}$.
In the case with large $k_\perp$
the TMD light-cone wave function is determined in term of the TMI one as:
\begin{equation}
\phi_+ (k^+,k_\perp, \zeta, \mu ) = \int_0^\infty d q^+
C_\perp(k^+, q^+,k_\perp, \zeta )
\Phi_+ (q^+,\mu)   + {\mathcal O}((k^2_\perp)^{-2}),
\end{equation}
where the function $C_\perp$ can be determined by perturbative QCD.
By power counting\cite{BFJMY} $C_\perp$ is proportional to $(k_\perp^2)^{-1}$.
When we consider the Fourier-transformed TMD light-cone wave function
into the impact $b$-space:
\begin{equation}
\phi_+(k^+, b, \zeta,\mu) = \int d^2 k_\perp e^{i \vec k_\perp \cdot \vec b}
       \phi_+(k^+,k_\perp, \zeta,\mu),
\end{equation}
$\phi_+(k^+, b, \zeta,\mu)$ can be related to the TMI one for small $b$ as:
\begin{equation}
\phi_+(k^+,b,\zeta, \mu ) = \int_0^\infty d q^+ C_B(k^+,q^+, b, \zeta,\mu)
 \Phi_+(q^+,\mu) +{\mathcal O}(b),
\end{equation}
where $C_B$ can also be calculated with perturbative
QCD, i.e., it does not contain any soft divergence. Hence the relation represents
a factorization. The leading order
result is $C_B^{(0)} =\delta(k^+ -q ^+)$.
A similar factorization between
TMD- and usual  parton distribution was proven in \cite{CS}.
We will determine the relation in Eq.(5) and Eq.(7)
up to order of $\alpha_s$ and show that the factorization in Eq.(7)
holds at one-loop level. In determination of these relations
we also derive the $\mu$-evolution equation of the TMI light-cone wave function
and the Collins-Soper equation of the TMD one.
\par
The relation in Eq.(5) is useful for constructing models of the TMD
light-cone wave function. The importance of the relation in the $b$-space
in Eq.(7) and the Collins-Soper equation is the following:
When the TMD factorization is formulated in the $b$-space,
the relation allows to use the TMI light-cone wave function, while
the Collins-Soper equation is used to resum large logarithms. Hence
it is possible to have relations between two factorization approaches
under certain conditions. It should be noted that other definitions
of a TMD light-cone wave function are possible. A different definition
is given in \cite{LiLiao} with a complicated structure of gauge links.
With this definition the relations in Eq.(5) and Eq.(7) can also be studied.
\par
The functions $C_\perp$ and $C_B$ are free from any soft
divergence, i.e., infrared- and collinear singularities, we can
use a partonic state instead of a B-meson state to determine them.
We use the partonic state $\vert b(m_b v +k_b), \bar
q(k_q)\rangle$, the momenta are given as $k_q^\mu =(k_q^+, k_q^-,
\vec k_{q\perp})$ and $k_b^\mu =(k_b^+, k_b^-, -\vec k_{q\perp})$.
These partons are on-shell, i.e., $k_q^2 =m_q^2$ and $v\cdot
k_b=0$ in HQET. The variable $k^+$ of the wave functions is from
$0$ to $\infty$. Actually, from the momentum conservation, it is
from $0$ to $P^+ =m_bv^+ +k_b^+ + k_q^+$. Under the limit $m_b\to
\infty$ we have $P^+ \to \infty$. If we set $P^+$ to be $\infty$
at the beginning, it will results in an ill-defined distribution
like $1/(k^+ -k_q^+)_+$ for $k^+$ going to $\infty$. Therefore we
will take a finite $P^+$ in the calculation and take the limit
$P^+ \to \infty$ in the final result.  We illustrate this in
detail for $\Phi_+$.
\par
\vskip20pt


\begin{figure}[hbt]
\begin{center}
\includegraphics[width=9cm]{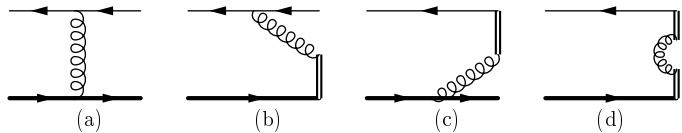}
\end{center}
\caption{Diagrams of one-loop contributions. Thick lines stand for
$b$-quark, double lines represent gauge links.  }
\label{Feynman-dg1}
\end{figure}
\par\vskip20pt
At leading order
$ \Phi_+ (k^+,\mu ) = \bar v(k_q)\gamma^+\gamma_5 u_v(k_b)
      \delta (k^+ -k_q^+)$.
At order of one loop, the contributions are from Feynman diagrams
shown in Fig.1 and Fig.2., except from
Fig.1d, Fig.2b and Fig.2e.
The contributions from Fig.2. are proportional
to the tree-level result.
We use dimensional regularization with
$d=4-\varepsilon$ for U.V. divergence and give gluons a small mass
$\lambda$ for infrared divergences. There are light-cone singularities
in diagrams where the gluon is attached to a gauge link. They
are cancelled
between different diagrams. To show this, we take Fig.1c. as an example.
The contribution from Fig.1c after the integration of the gluon momentum is:
\begin{eqnarray}
\Phi_+ (k^+,\mu)\vert_{1c}
   =  -\frac{ 2\alpha_s}{3\pi^2 }
   \bar v(k_q)  \gamma^+\gamma_5 u(k_b)  \theta(k^+ -k_q^+)
 (4\pi \mu^2)^{\varepsilon/2} \Gamma(\varepsilon/2)
     \frac{(\lambda^2+\zeta^2_v (1-x)^2)^{-\varepsilon/2}}{k_q^+-k^+},
\end{eqnarray}
where $x=k_q^+/k^+$ and $\zeta_v^2 = 4(v^- k_q^+)^2$.
This contribution has the light-cone singularity at $k^+ =k_q^+$.
The contribution from Fig.2c reads:
\begin{equation}
\Phi_+ (k^+,\mu)\vert_{2c} = -\delta (k^+-k_q^+)
    \bar v(k_q) \gamma^+ \gamma_5 u(k_b) \frac{2\alpha_s}{3\pi}
     (4\pi \mu^2)^{\varepsilon/2} \Gamma(\varepsilon/2)
     \int_0 ^\infty d p^+\frac{(\lambda^2+\zeta^2_v (p^+/k_q^+)^2)^{-\varepsilon/2}}{p^+}.
\end{equation}
The integral is divergent for $p^+ \to\infty$ and $p^+ \to 0$.
The divergence at $p^+ =0$ is the light-cone singularity and will be cancelled
with that from Fig.1c. If we set $P^+$ to be $\infty$ at the beginning,
the sum of these two contributions integrated with a test function $f(k^+)$
is proportional to
\begin{equation}
\int_{k_q^+}^{\infty} d k^+
  \frac{ f(k^+) -f(k_q^+)}{k^+-k_q^+} \ln \frac{(\mu k_q^+)^2}{\zeta_v^2(k^+ -k_q^+)^2}.
\end{equation}
This integral is divergent, because the divergence at $p^+ \to \infty$ in Eq.(9)
is still there.
The divergence has a geometrical reason\cite{LN}.
In HQET the $b$-quark field $h$ in Eq.(1) can be represented as a gauge link along
the direction $v$ and forms a Wilson line combining the gauge link $L_n$.
This Wilson line has a cusp singularity at the origin where the two gauge links
join\cite{Pol}. This divergence needs to be renormalized.
With a finite
$P^+$ the sum becomes proportional to the expression
instead of the integral in Eq.(10):
\begin{equation}
\int_{k_q^+}^{P^+} d k^+
    \frac{f(k^+)-f(k_q^+)}{k^+-k_q^+} \ln \frac{(\mu k_q^+)^2}{\zeta_v^2(k^+ -k_q^+)^2}
 -\frac{f(k_q^+)}{4}\left [ \ln^2 \frac{(\mu k_q^+)^2}{\zeta_v^2 (P^+ -k_q^+)^2 }
        +\xi(2) \right ].
\end{equation}
One can easily check that the expression is finite under the limit $P^+\to \infty$.
Evaluating all diagrams we have the result at one loop with $x_P =P^+/k_q^+$:
\begin{eqnarray}
\Phi_+ (k^+,\mu ) &=&\bar v(k_q) \gamma^+ \gamma_5 u(k_b) \delta (k^+- k_q^+)
         + \Phi_+(k^+,\mu)_{1a}
\nonumber\\
 && +  \frac{2\alpha_s}{3\pi }
 \bar v(k_q) \gamma^+ \gamma_5 u(k_b) \Big \{
  \delta (k^+-k_q^+) \Big [ \frac{1}{4}\ln\frac{\mu^2}{m_q^2}
     -\ln\frac{\lambda^2}{m_q^2} -1  -\frac{1}{4} \ln^2 \frac{\mu^2}{\zeta_v^2 (x_P -1)^2}
    -\frac{\pi^2}{24} \Big ]
\nonumber\\
 &&  +\left (\frac{k^+ \theta(k_q^+-k^+)}{k_q^+(k_q^+ -k^+)} \ln\frac{\mu^2}{m_q^2(1-x)^2} \right )_+
     +\left (\frac{\theta(k^+-k_q^+)}{k_q^+-k^+} \ln \frac{\zeta_v^2(1-x)^2}{\mu^2}
   \right )_+  \Big \},
\end{eqnarray}
the contribution from Fig.1a is U.V. finite and is not needed for deriving our final
results, as explained later. The above expression is a distribution for $ 0 < k^+ < P^+$.
From the above result
one derives the renormalization evolution under the limit $P^+ \to \infty$:
\begin{eqnarray}
\mu \frac{\partial}{\partial \mu} \phi_+ (k^+,\mu ) &=& \int_0^{\infty}  dq^+
   \gamma_+ ({k^+}, {q^+},\mu ) \phi_+ (q^+, \mu),
\nonumber\\
\gamma_+ (k^+, q^+, \mu ) &=& \frac{4\alpha_s}{3\pi}\left \{
  \big(\frac{5}{4}-\ln\frac{\mu}{2v^- k^+ } \big)\delta (k^+ - q^+)
   + \left [ \frac{k^+ \theta(q^+- k^+ )}{ q^+ ( q^+ - k^+)}
     +\frac{\theta(k^+ - q^+)}{ k^+  - q^+} \right ]_+ \right \} .
\end{eqnarray}
In deriving this one should be careful with the plus description acting on different
distribution variables. The plus prescription above is for $q^+$. This result
is in agreement with that in \cite{LN} by noting the fact that the wave function
defined in \cite{LN} is with the decay constant in HQET.
From our explicit result it is observed that the wave function can not be normalized
under the limit $P^+ \to \infty$ as first observed
in \cite{LCB} and in other studies\cite{norm}.
\par
To determine the mentioned relations we need to calculate the wave function
$\phi_+$ at one-loop order. All diagrams in Fig.1 and Fig.2 give
contributions. The calculation is straightforward in the momentum space.
The light-cone singularity is regularized by a small but finite $u^+$.
The detailed calculation and result will be presented elsewhere and we will
only give final results here.
To determine the function $C_\perp$ we only need to calculate
Fig.1b, Fig.1c and Fig.1d. The contribution from Fig.1a is proportional
to $1/(k_\perp^2)^2$ for large $k_\perp$ and will not contribute to $C_\perp$.
The function is determined by taking the limit of large $k_\perp$ and then $P^+\to \infty$.
We obtain:
\begin{eqnarray}
C_\perp (k^+,q^+,k_\perp, \zeta ) = \frac{2\alpha_s}{3\pi}\cdot \frac{1}{k^2_\perp}\left [
 \left (  \frac{k^+ \theta(q^+-k^+)}{q^+(q^+-k^+)}
     + \frac{\theta(k^+-q^+)}{k^+-q^+} \right )_+
 +\delta (k^+ -q^+)
  (\ln \frac{\zeta^2}{k^2_\perp} -1)
 \right ].
\end{eqnarray}
\par\vskip20pt

\begin{figure}[hbt]
\begin{center}
\includegraphics[width=9cm]{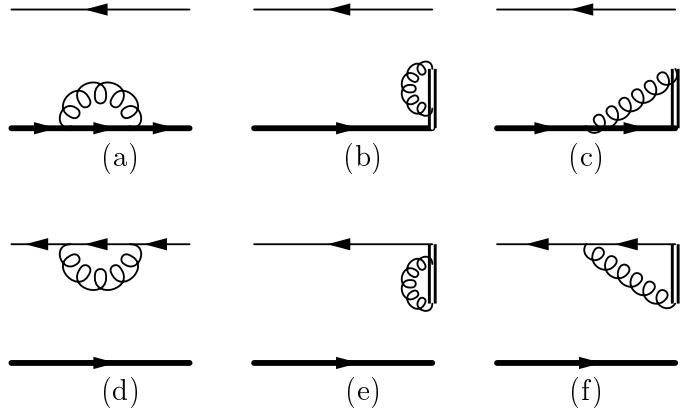}
\end{center}
\caption{The self-energy corrections. }
\label{Feynman-dg1}
\end{figure}
\par\vskip20pt
\par
To determine the function $C_B$ we do not need to calculate the contribution
from Fig.1a, again. The reason is that the contribution
$\phi_+(k^+, k_\perp, \zeta, \mu)_{1a}$
is U.V. finite when integrated over $k_\perp$. That means that
$\phi_+(k^+, b, \zeta, \mu)_{1a}=\phi_+(k^+, b=0, \zeta, \mu)_{1a} +{\mathcal O}(b)$
for $b\to 0$. For small $b$ we have:
\begin{eqnarray}
\phi_+ (k^+,b, \zeta,\mu) &=&\bar v(k_q) \gamma^+ \gamma_5 u(k_b) \Big \{
\delta (k^+ -k_q^+)
 +\frac{2\alpha_s}{3\pi }
    \Big [ \delta (k^+-k_q^+)
     \big ( -\frac{1}{4} \ln^2 \zeta^2 \tilde b^2 -\frac{1}{2}
      \ln \mu^2 \tilde b^2 \ln\frac{\zeta^2}{\zeta_v^2}
\nonumber\\
 &&  -\frac{1}{4}\ln^2 \zeta_v ^2 \tilde b^2 (x_P-1)^2
  +2 \ln \mu^2 \tilde b^2  +\frac{1}{2}\ln\frac{\zeta^2}{\mu^2}
         -\ln\frac{\lambda^2}{m_q^2} +\frac{1}{4}\ln\frac{\mu^2}{m_q^2}
         -2 -\frac{\pi^2}{3} \big )
\nonumber\\
&&  +  \Big  (\frac{\theta(k^+ - k_q^+ )}{k_q^+ -k^+}
           \ln(\tilde b^2 \zeta^2_v(1-x)^2)
- \ln (\tilde b^2  m_q^2(1-x)^2)
                  \frac{k^+ \theta (k_q^+ - k^+ )}{k_q^+ (k_q^+ -k^+)}\Big )_+
          \Big ]\Big \}
\nonumber\\
         &&  + \phi_+(k^+, b=0, \zeta, \mu)_{1a} +{\mathcal O}(b),
\end{eqnarray}
with $\tilde b^2 = b^2 e^{2\gamma}/4$. This result is a distribution
for $ 0< k^+ < P^+$.
From the above result one can derive the Collins-Soper equation:
\begin{eqnarray}
\zeta \frac{\partial}{\partial \zeta} \phi_+(k^+,b, \zeta,\mu) =
   \left [-\frac{4\alpha_s}{3\pi} \ln\frac{\zeta^2 b^2 e^{2\gamma-1}}{4}
       -\frac{2\alpha_s}{3\pi} \ln\frac{\mu^2e}{\zeta^2} \right ]
       \phi_+(k^+,b,\zeta,\mu).
\end{eqnarray}
The first factor is the famous factor $K+G$\cite{CS,CSS},
the last factor comes because we used HQET for the heavy quark.
\par
Comparing the result in Eq.(12) and Eq.(15) and noting the fact that
$\phi_+(k^+, b=0, \zeta, \mu)_{1a}$ is just
$\Phi_+ (k^+,\mu)_{1a}$, we can derive
the function $C_B$ under the limit $P^+ \to \infty$:
\begin{eqnarray}
C_B(k^+, q^+,b,\zeta ) &=&\delta(k^+ -q^+) +  \frac{2\alpha_s}{3\pi}\Big \{
   \ln(\mu^2 \tilde b^2 )\left  [  -
                  \frac{k^+\theta (q^+ - k^+)}{ q^+ (q^+- k^+)}
 +  \frac{\theta(k^+ - q^+)}{q^+ - k^+}
           \right ]_+
\nonumber\\
   && +  \delta (k^+ - q^+ )
     \Big [ \frac{1}{4} \ln^2 \frac{\mu^2}{\zeta^2}
            -\frac{1}{2} \ln^2 (\zeta^2 \tilde b^2)
 + \ln (\mu^2 \tilde b^2)  +\frac{1}{2}\ln\frac{\zeta^2}{\mu^2}
         -1 -\frac{7\pi^2}{24}   \Big ] \Big \} ,
\nonumber\\
\end{eqnarray}
which does not contain any soft divergence and does not depend on $v$.
In the above the plus prescription is for $q^+$.
\par
To summarize: We proposed the definition in Eq.(3) for the TMD light-cone wave function
of a $B$-meson. The definition does not contain the light-cone singularity and
can be used for performing TMD factorization in a consistent way.
Two relations between TMD- and TMI light-cone wave function are found.
One is that the TMD light-cone wave function with large $k_\perp$ is determined
by the TMI one.
This relation will be important for constructing models of the TMD
light-cone wave function.
Another one
is the factorization relation between the TMI light-cone
wave function and TMD one in the impact $b$ space with the small $b$.
In studying these relations we also obtained the renormalization
evolution of the TMI light-cone wave function
and the Collins-Soper equation of the TMD one. The equation and the relation
in the impact space are important. When TMD factorization is formulated
in the impact space, the relation allows us to use the TMI light-cone wave function
and the equation allows us to resum large logarithms. These issues  will be discussed
in another publication in the near future.
\vskip 5mm
\par\noindent
{\bf\large Acknowledgments}
\par
The authors would like to thank Prof. X.D. Ji, H.-n. Li, M. Neubert and C.J. Zhu for discussions
and communications.
This work is supported by National Nature
Science Foundation of P.R. China.
\par\vskip20pt


\end{document}